\documentclass[epj]{svjour}
\usepackage{graphics}
\usepackage{graphicx}
\usepackage{xcolor}
\usepackage[symbol]{footmisc}
\newcommand{\kaonangle}{$\cos\theta_\mathrm{CM}^{K}$}
\begin{document}
\title{$K^+\Lambda(1520)$ photoproduction at forward angles near threshold with the BGOOD experiment}
\author{%
	E.O.~Rosanowski\inst{1,}\thanks{Corresponding author:  s6emrosa@uni-bonn.de}%
	\and T.C.~Jude\inst{1,}\thanks{Corresponding author:  jude@physik.uni-bonn.de}%
	\and S.~Alef\inst{1,}
		\and A.J. Clara~Figueiredo\inst{1}
		\and{D.D Burdeinyi}\inst{2}
	\and P.L.~Cole\inst{3}
	\and R.~Di Salvo\inst{4}%
	\and{D.~Elsner}\inst{1}
	\and A.~Fantini\inst{4,5}%
	\and O.~Freyermuth\inst{1}%
	\and F.~Frommberger\inst{1}%
	\and V.B~Ganenko\inst{2}
	\and F.~Ghio\inst{6,7}%
	\and J.~Gro\ss \inst{1}%
	\and K.~Kohl\inst{1}
	\and P.~Levi Sandri\inst{8}%
	\and G.~Mandaglio\inst{9,10}%
	\and R.~Messi\inst{4,5}
	\and D.~Moricciani\inst{4,8}\thanks{Deceased}
	\and P.~Pedroni\inst{11}%
	\and B.-E.~Reitz\inst{1}
	\and M.~Romaniuk\inst{12}
	\and G.~Scheluchin\inst{1}
	\and H.~Schmieden\inst{1}%
	\and A. Sonnenschein\inst{1}
}
\institute{%
	Rheinische Friedrich-Wilhelms-Universit\"at Bonn, Physikalisches Institut, Nu\ss allee 12, 53115 Bonn, Germany
	\and National Science Center “Kharkov Institute of Physics and Technology”, UA61108 Kharkov, Ukraine
		\and Lamar University, Department of Physics, 77710, Beaumont, Texas,  USA
		\and INFN Roma ``Tor Vergata", Via della Ricerca Scientifica 1, 00133, Rome, Italy
			\and Universit\`a di Roma ``Tor Vergata'', Dipartimento di Fisica, Via della Ricerca Scientifica 1, 00133, Rome, Italy
				\and INFN sezione di Roma La Sapienza, P.le Aldo Moro 2, 00185, Rome, Italy 
					\and Istituto Superiore di Sanit\`a, Viale Regina Elena 299, 00161, Rome, Italy 
		\and INFN - Laboratori Nazionali di Frascati, Via E. Fermi 54, 00044, Frascati, Italy
		\and INFN sezione Catania, 95129, Catania, Italy
		\and Universit\`a degli Studi di Messina, Dipartimento MIFT,  Via F. S. D'Alcontres 31, 98166, Messina, Italy
	\and INFN sezione di Pavia, Via Agostino Bassi, 6 - 27100 Pavia, Italy
	\and Institute for Nuclear Research of NASU, 03028, Kyiv, Ukraine
}

%
%
%
\date{Received: date / Revised version: date}
%
\abstract{The differential cross section for $\gamma p\rightarrow K^+\Lambda(1520)$ was measured from threshold to a centre-of-mass energy of 2090\,MeV at forward angles at the BGOOD experiment.  The high statistical precision and resolution in centre-of-mass energy and angle allows a detailed characterisation of this low-momentum transfer kinematic region.  The data agree with a previous LEPS measurement and support effective Lagrangian models that indicate that the contact term dominates the cross section near threshold.
} 
\maketitle

\section{Introduction}
\label{sec:introduction}

Strangeness photoproduction is a crucial tool to understand baryon resonance spectra and the underlying degrees of freedom afforded in the low energy, non-perturbative regime of QCD.  $K\Lambda^*$ and $K\Sigma^*$ photoproduction has frequently been used to probe the third resonance region in attempts to characterise known higher mass $N^*$ and $\Delta$ states and search for so called \textit{missing resonances} which have been predicted by constituent quark models~\cite{capstick86,capstick92,capstick94,riska01,klempt10,klempt12}, lattice QCD calculations~\cite{edwards11}, and Dyson-Schwinger equations of QCD~\cite{roberts11} but not observed experimentally in $N\pi$ final states~\cite{capstick00,loering01}.  $K\Lambda^*$ channels are unique in that the isoscalar $\Lambda$ acts as an isospin filter, permitting only  intermediate $s$-channel $N^*$ resonances, and in the case of the ground state $\Lambda$, the self-analysing weak decay enables easier access to baryon-recoil polarisation observables.  

The $\Lambda^*$ excitation spectrum however is itself poorly understood compared to the non-strange baryon sector, mainly due to a scarcity of experimental data~\cite{pdg20}.  Very few measurements on $S=-1$ resonances (both $\Lambda^*$ and $\Sigma^*$ states)  have been made over the last 20 years.  The exception to this are measurements of the final states $K^+\Sigma^0\pi^0$ and $K^+\Sigma^\mp\pi^\pm$~\cite{lu13,moriya13,moriya13b,niiyama08,lambda1405paper} in both electro- and photoproduction, which have mainly focussed on decays of the $\Lambda(1405)$.  Studies of the $\Lambda(1405)$ are largely driven by its exotic nature and it is now considered the archetypal molecular-like hadron in the $uds$ quark sector~\cite{guo18}.  The large mass difference to the spin-orbit partner, the $\Lambda(1520)$ is also hard to explain within a CQM, and given  the strong evidence of the molecular nature of the $\Lambda(1405)$, it is natural to question the structure of the $\Lambda(1520)$.  Refs.~\cite{kolomeitsev04,sarkar05,sarkar05b}, for example employed a chiral coupled channel model  which described the $\Lambda(1520)$ as a quasi-bound state of $\pi\Sigma(1385)$ and $K\Xi(1530)$. 

Previous measurements of $K^+\Lambda(1520)$ photoproduction near threshold were made at the LEPS~\cite{kohri10}, CLAS~\cite{moriya13b,shrestha21} and SAPHIR~\cite{wieland11} experiments\footnote{There are also earlier measurements made at 11\,GeV at the SLAC spectrometer~\cite{boyarski71} and between 2.8 and 4.8\,GeV from the LAMP2 collaboration~\cite{barber80}.}, which have been supported by models employing effective Lagrangians~\cite{he12,nam05,nam07,nam10,toki08,xie10,xie14,wie21} and models including additional Regge trajectories to describe higher energies~\cite{nam10b,wang14,he14,yu17}.  These analyses generally agree that near threshold the contact term dominates, with $t$-channel $K$ exchange important at higher energies.  Contributions from $t$-channel $K^*$ exchange, $u$-channel $\Lambda$ exchange and varying $s$-channel $N^*$ contributions however vary significantly, largely due to the unknown $\bar{K}^*N\Lambda^*$ coupling and choice of hadronic form factors.  The model of He and Chen~\cite{he12} for example, was used to determine that  the contact term dominates close to threshold, with $K^*$ $t$-channel contributions at higher centre-of-mass energies.  Two $N^*$ resonances with spins 3/2 and 5/2 were also required with masses close to 2.1\,GeV for an adequate reproduction of the data.
Similarly, the model of Xie, Wang and Nieves supported the contribution of an $N^*(2120)$ with  $J^P = 3/2^-$~\cite{xie10,xie14}.  This corroborates the results of a coupled channel model of Ramos and Oset~\cite{ramos13}, which suggested a dynamically generated state at 2\,GeV with $J^P = 3/2^-$  plays a dominant role in $K^0\Sigma$ photoproduction~\cite{ewald12,k0paper}.  The analyses by Wei et al.~\cite{wie21} employed the first effective Lagrangian model to make a global fit to differential cross section and beam asymmetry data, with the strong motivation  that  previous  effective Lagrangian models were unable to describe LEPS beam asymmetry measurements~\cite{kohri10}.  Two fits were made, either including an $N(2050)5/2^-$ or $N(2120)3/2^-$ resonance (denoted Fit A and B respectively), both giving satisfactory results.  Whilst the contact term and $t$-channel $K$ exchange were dominant in both models, contributions from $K^*$ and $u$-channel $\Lambda$ exchange varied markedly.

%
%

If the $\Lambda(1520)$ is a  candidate for a molecular-like state, it is of interest to determine the dependency that momentum transfer has upon the differential cross section.  It would be expected that a loosely bound system breaks up under large momentum transfer, which is observed for example, in coherent reactions off the deuteron and light nuclei (see for example, Ref.~\cite{krusche11}).  Accessing forward $K^+$ angles close to threshold minimises the momentum transfer (minimising the Mandelstam variable, $t$) to the $\Lambda(1520)$, and an enhancement or structure may be expected if a meson-baryon type ``resonating structure" contributes, for example, a molecular-type configuration.  The BGOOD photoproduction experiment~\cite{technicalpaper} at the ELSA facility~\cite{hillert06,hillert17} at the University of Bonn is ideally suited to measure associated strangeness in this kinematic regime.  The $K^+$ can be identified at forward angles in the \textit{Forward Spectrometer}, while the hyperon decays almost isotropically,  the decay products of which are identified in the central region with the \textit{BGO Rugby Ball} electromagnetic calorimeter.  The BGOOD collaboration is pursuing a programme of strangeness photoproduction measurements under low-momentum transfer conditions~\cite{lambda1405paper,klambdapaper,ksigmapaper} which have not been achievable with high statistical precision at other facilities.  An example of the importance of studying this region includes the forward differential cross section for $K^+\Sigma^0$, where a cusp-like structure was observed at the $K^+\Sigma(1385)$ threshold~\cite{ksigmapaper,INPCC22Conf}.  The measurement of $K^+\Lambda(1520)$ photoproduction close to threshold at forward angles is therefore the mandatory next step in order to understand the role of potential molecular-like structure in the strangeness sector.       
	
\section{Experimental setup and analysis procedure}
\label{sec:exp}

\begin{figure}[h]
		\includegraphics[trim={0cm 0cm 0cm 0cm},clip,width=0.5\textwidth]{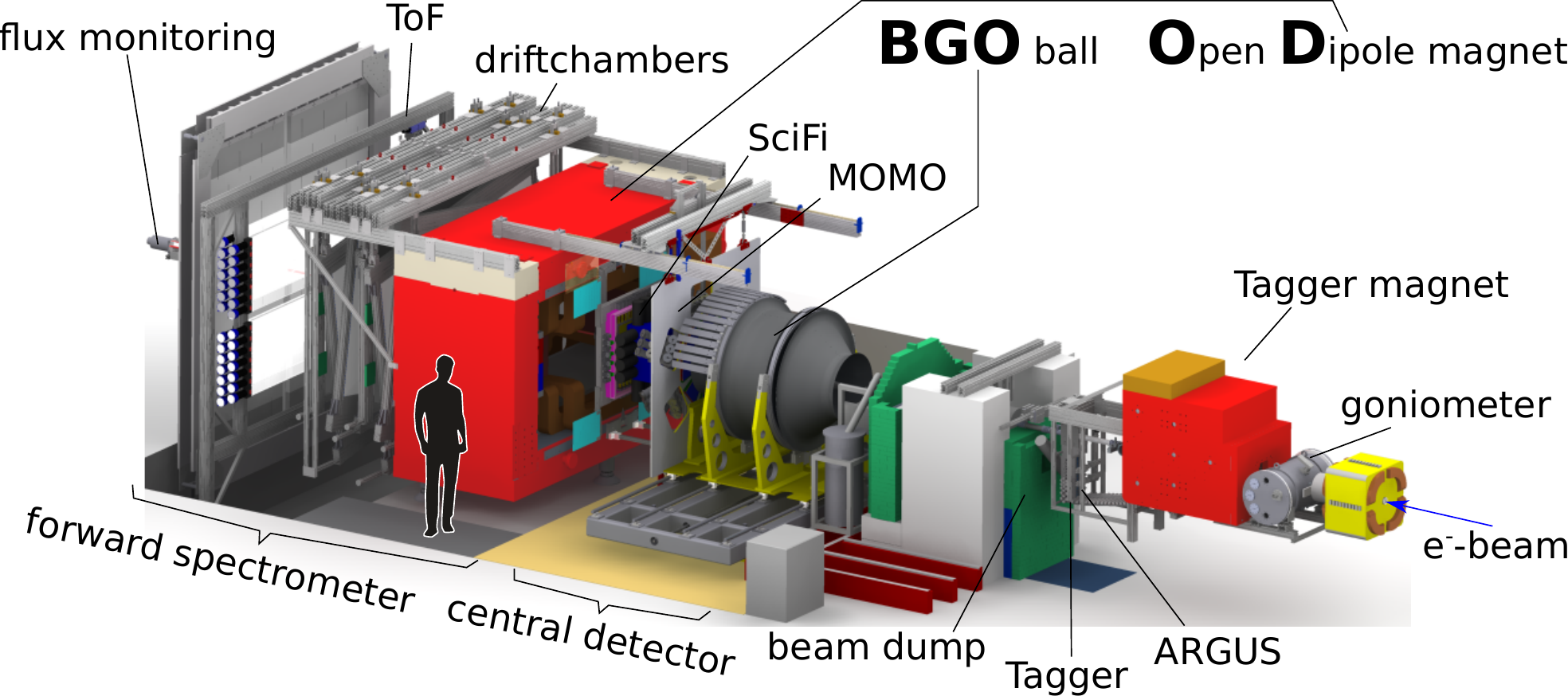}	
	\caption{The BGOOD experimental setup.  Figure originally shown in Ref.~\cite{technicalpaper}.}
	\label{fig:BGOODsetup}
\end{figure}

BGOOD, shown in Fig.~\ref{fig:BGOODsetup}, is comprised of two main parts: a central calorimeter region, ideal for neutral meson identification, and a magnetic \textit{Forward Spectrometer} for charged particle identification and momentum reconstruction (for a detailed description see Ref.~\cite{technicalpaper}).  The \textit{BGO Rugby Ball} is the main detector over the central region, covering laboratory polar angles 25 to 155$^\circ$.  The detector is comprised of 480 BGO crystals for the reconstruction of photon momenta via electromagnetic showers in the crystals.  The separate time readout per crystal enables a clean separation and identification of neutral meson decays.  Between the BGO Rugby Ball and the target are the \textit{Plastic Scintillating Barrel} for charged particle identification via $\Delta E-E$ techniques and the  \textit {MWPC} for charged particle tracking and vertex reconstruction.

The Forward Spectrometer covers a laboratory polar angle 1-12$^\circ$. The tracking detectors, \textit{MOMO} and \textit{SciFi} are used to track charged particles from the target.  Downstream of these is the \textit{Open Dipole Magnet} operating at an integrated field strength of 0.216\,Tm.  A series of eight double layered \textit{Drift Chambers} track charged particles after curvature of their trajectories in the magnetic field and are used to determine particle momenta with a resolution of approximately 6\,\%.\footnote{The resolution improves to 3\,\% if the Open Dipole Magnet is operating at the maximum integrated field strength of 0.432\,Tm.}  Three \textit{Time of Flight (ToF) Walls} downstream of the drift chambers determine particle $\beta$ and are used in combination with the measured momentum for particle identification via  mass determination.
Track reconstruction in the Forward Spectrometer is described in Ref.~\cite{technicalpaper}. 

The small intermediate region between the central region and the Forward Spectrometer is covered by \textit{SciRi}, which consists of three concentric rings, each with 32 plastic scintillators for charged particle detection. 

The presented results are from data taken over two periods.  The first dataset was taken in 2017 and used a 6\,cm long liquid hydrogen target with a 3.2\,GeV electron beam over 22 days, resulting in an integrated photon flux of  $1.11\times 10^{12}$ over the photon energy range 1650 to 1850\,MeV (the region of the presented data).  The second dataset taken in 2018 used an 11\,cm long liquid hydrogen target with a 2.8\,GeV electron beam over 18 days, resulting in an integrated photon flux of  $7.28\times 10^{11}$ over the same photon energy range.  Four different hardware triggers were implemented to ensure the data can be used for a variety of measurements (see Refs.~\cite{technicalpaper,klambdapaper}), however the results presented here required a single trigger combination, which was an incident photon registered in the photon tagger and a minimum energy deposition of 150\,MeV in the BGO Rugby Ball.

$K^+$ were identified in the Forward Spectrometer by reconstructing their invariant mass from  momentum and $\beta$ measurements.  $K^+$ candidates were selected over a $2\sigma$ range of the reconstructed mass.  This increased linearly with $K^+$ momentum, ranging from 45\,MeV/c$^2$ to 73\,MeV/c$^2$ at momenta 450 and 1000\,MeV/c respectively.

Two different topologies of the final state were required for consistency checks and to constrain systematic uncertainties.  The first required only a $K^+$ candidate in the Forward Spectrometer, and the second required an additional $\pi^0$ to be identified in the BGO Rugby Ball via the decay, $\pi^0\rightarrow \gamma \gamma$.  $\pi^0$ candidates were selected where the   $\gamma\gamma$ invariant mass was within 30\,MeV of the nominal $\pi^0$ mass (corresponding to $2\sigma$).  The requirement of the $\pi^0$ reduces the dataset to candidate events where the decays, $K^+(\Lambda(1520)\rightarrow \Sigma^0\pi^0)$ and $K^+(\Lambda(1520)\rightarrow \Lambda\pi^0\pi^0)$ are observed.

\begin{figure}[h]
	\includegraphics[trim={0cm 2.5cm 20cm 0cm},clip,width=0.5\textwidth]{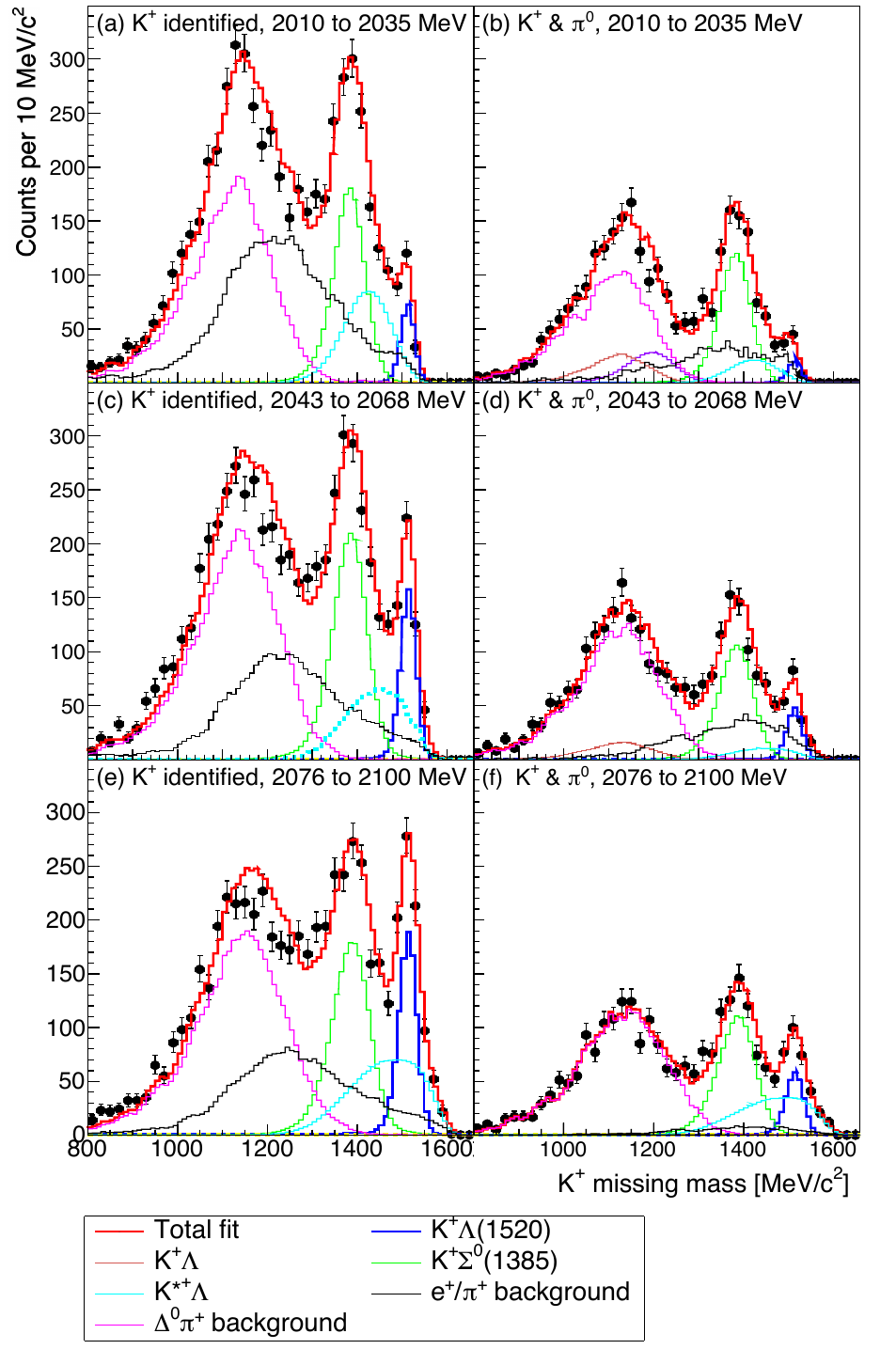}
	\caption{Examples from the 2018 dataset of the missing mass recoiling from the $K^+$ for the two topologies  for different $W$ intervals labelled inset.  Left panels are for when only a forward $K^+$ candidate is required and right panels are for when an additional $\pi^0$ candidate is required.  Each $W$ interval corresponds to two tagger channels.  The fitted distributions from simulated data (and real data for the $e^+\pi^+$ background described in the text) are listed in the legend.}\label{fig:MMFit}
\end{figure}

The event yields contained background from hadronic reactions, where either the forward $K^+$ was misidentified from $\pi^+$ or $e^+$, or other associated strangeness reactions of $K^+Y$ and $K^+ Y\pi$.  This background can be observed in histograms of the missing mass recoiling from the $K^+$ in the final state, examples of which are shown in Fig.~\ref{fig:MMFit} for the two different topologies (a $K^+$ candidate with or without an additional $\pi^0$ candidate).  To extract the $K^+\Lambda(1520)$ yields, these were fitted with distributions from simulated associated strangeness channels.  This was achieved using the BGOOD GEANT4~\cite{geant4} simulation, including all spatial, energy and time resolutions, magnetic fields and hardware efficiencies (see Ref.~\cite{technicalpaper} for details).  Event distributions were based on measured data where available or phase distributions, however the small $W$ and \kaonangle{} range per interval made differences in distributions negligible.  Signal events ($K^+\Lambda(1520)$) and background from $K^+\Lambda$, $K^+\Sigma^0$, $K^+\Sigma^0(1385)$ and $K^{*+}\Lambda$ were simulated.  $K^+\Lambda(1405)$ was not included due to the mass degeneracy to the $\Sigma(1385)$.  From previous studies~\cite{klambdapaper}, it was realised that $\Delta \pi^+$ was required as it gave substantial background contribution from $K^+$-$\pi^+$ misidentification, however this only contributes at lower missing mass than the $K^+\Lambda(1520)$ signal.  An additional background distribution from misidentifying either  $\pi^+$ or $e^+$ as $K^+$ was determined by repeating the analysis but requiring a negatively charged track (a technique first used in Refs.~\cite{klambdapaper,ksigmapaper}).  This resulted in $\pi^-$ and $e^-$ distributions which appeared (by eye) identical to their positively charged counterparts.  The measured data extends only to the $K^*\Sigma^0$ threshold as the missing mass distribution from this background channel was too close to the $K^+\Lambda(1520)$ distribution for a reliable separation of signal and background.\footnote{Preliminary measurements of $K^*Y$ channels at BGOOD have been made and are under analysis for future publications.  Once complete, it  is the intention to subtract the $K^*\Sigma^0$ contributions and extend the measured $K^+\Lambda(1520)$ $W$ range.}  

\begin{figure}[h]
	\includegraphics[width=0.5\textwidth]{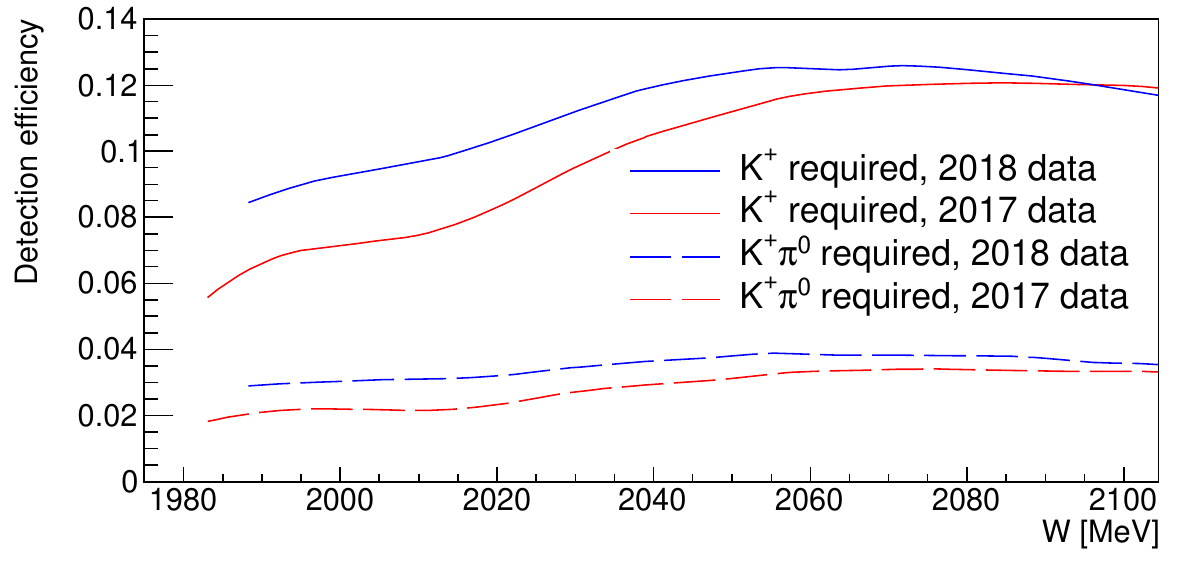}
	\caption{The detection efficiency for \kaonangle{}$> 0.9$ for the two different datasets and either requiring only a forward $K^+$ or an additional $\pi^0$ in the BGO Rugby Ball.}\label{fig:deteff}
\end{figure}

The detection efficiency was determined by simulating $K^+\Lambda(1520)$ events assuming a phase space distribution, including the known $\Lambda(1520)$ decay branching ratios.  Shown in Fig.~\ref{fig:deteff}, the detection efficiency when only requiring a $K^+$ has a maximum of approximately 12\,\%, whereas the requirement of an additional $\pi^0$ lowers the efficiency to approximately 3\,\%.  The small differences between the two datasets is due to the different target lengths and the efficiency of hardware triggers.

\begin{figure}[h]
	\includegraphics[width=0.5\textwidth]{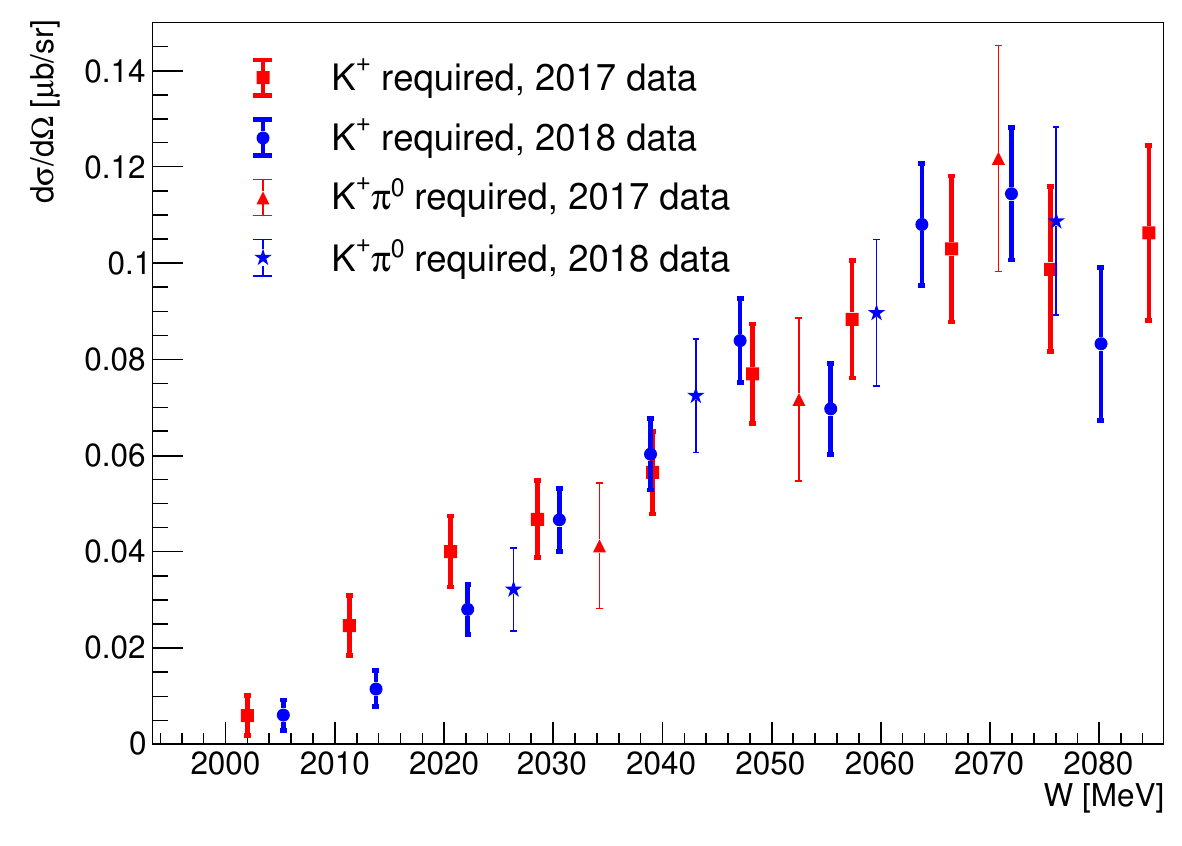}
	\caption{The differential cross section for \kaonangle{}$> 0.9$ for the two different datasets and either requiring only a forward $K^+$ or an additional $\pi^0$ in the BGO Rugby Ball (labelled inset).  The intervals in $W$ is either every photon tagger or every two photon tagger channels when requiring just the $K^+$ or the $K^+\pi^0$ combination respectively.}\label{fig:diagcs}
\end{figure}

Most of the systematic uncertainties listed in Table~\ref{table:syserror} were calculated for previous publications using the same datasets~\cite{klambdapaper}.  To determine the systematic uncertainty of the yield extraction from the fitting to the $K^+$ missing mass spectra, the differential cross section for the two datasets and the two different selection criteria were determined (shown in Fig.~\ref{fig:diagcs}).  An uncertainty of 5\,\% was estimated by comparing the measurements.  The total systematic uncertainty, when summed in quadrature is 9.3\,\%.

\begin{table}[h]
	\centering
	\begin{tabular}{l c}
		\hline\hline
		Source & \% error \\
		\hline
		Beam spot alignment & 4.0 \\
		Photon flux & 4.0\\
		$K^+$ selection & 2.0 \\
		SciFi efficiency & 3.0 \\
		Target wall contribution & 2.0 \\
		Track time selection & 2.0 \\
		Target length & 0.9 \\
		ToF wall efficiency & 1.5 \\
		MOMO efficiency & 1.0 \\
		Drift chamber efficiency & 1.0 \\
		Beam energy calibration & 1.0 \\
		Modelling of hardware triggers & 1.0 \\
		Forward track geometric selection & 1.0 \\
		Fitting $K^+$ missing mass spectra & 5.0 \\
		\hline
		Summed in quadrature & 9.3 \\
		\hline\hline
	\end{tabular}
	\caption{Systematic uncertainty contributions.}
	\label{table:syserror}
\end{table}

The results presented in the next section only required the $K^+$ detection (no $\pi^0$) to improve statistical precision and is an error weighted combination of both datasets.

\section{Results and interpretation}\label{sec:results}

\begin{figure}[h]
	\includegraphics[width=0.5\textwidth]{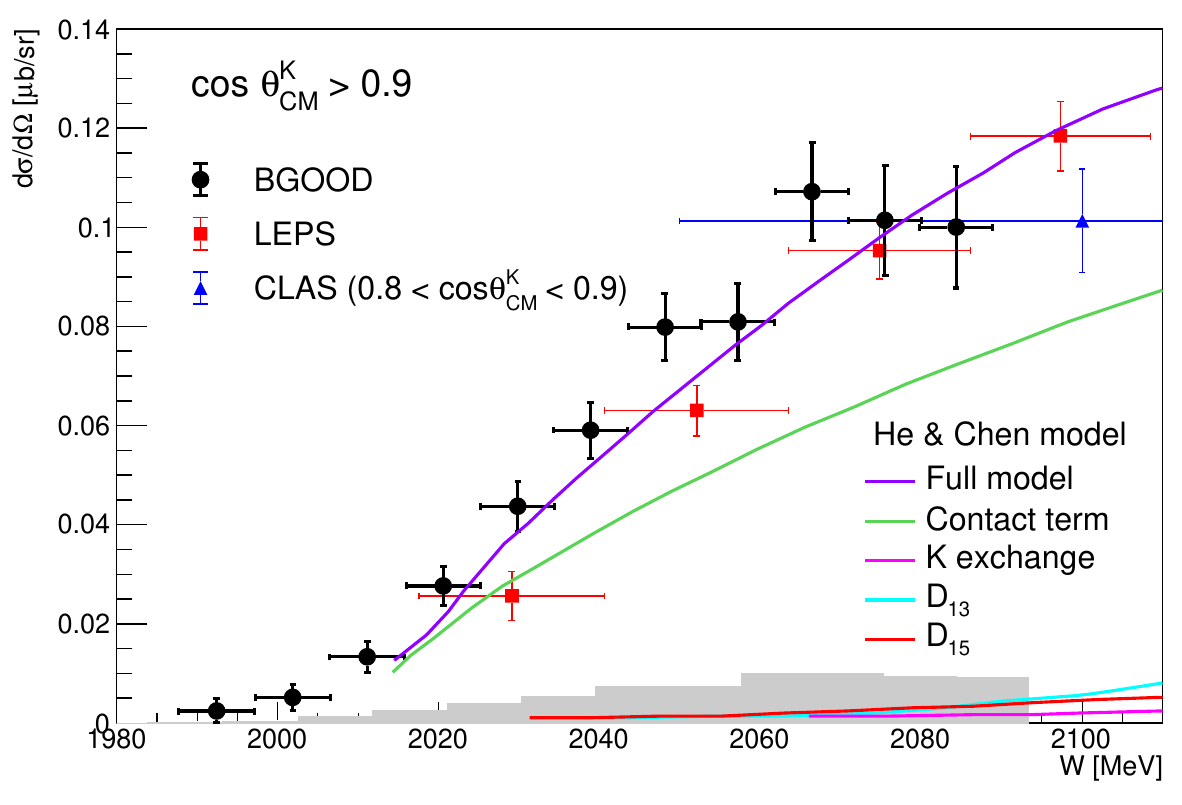}
	\caption{The differential cross section for \kaonangle{}$> 0.9$ for the combined BGOOD data set requiring only a forward $K^+$ (filled black circles with systematic errors indicated as the filled grey area on the abscissa).  Previous LEPS~\cite{kohri10} and CLAS data~\cite{moriya13b} (at the more backward \kaonangle{} of 0.8 to 0.9) are shown as filled red squares and filled blue triangles respectively.  For all data, vertical bars indicate statistical error and horizontal bars indicate the range of the data point. The results from the effective Lagrangian models of Xie and Nieves~\cite{xie10},  He and Chen~\cite{he12} and Wei et. al.~\cite{wie21} are superimposed and labelled inset.}\label{fig:cs}
\end{figure}

The differential cross section for \kaonangle{} $> 0.9$ is shown in Fig.~\ref{fig:cs} (filled black circles).  The data provide the first measurements with high precision close to threshold, achieving a resolution in $W$ of approximately 10\,MeV which is 2.5 times finer than previous LEPS data~\cite{kohri10} with similar statistical precision.   Superimposed are the effective Lagrangian models of He and Chen~\cite{he12}, Xie and Nieves~\cite{xie10,xie14} and Wei et. al~\cite{wie21} which were fitted to previous LEPS and CLAS data and give good agreement to the BGOOD data.  The He and Chen model, for example has a reduced $\chi^2$ of 1.96 when only including the statistical errors, however when combining the systematic and statistical errors in quadrature, a reduced $\chi^2$ of 1.04 is achieved.  The model of Wie et al.~used a global fit to differential cross section data and beam asymmetry measurements for both Fits A and B (including either $N(2060)5/2^-$ or $N(2120)3/2^-$ respectively), with both giving a reasonable agreement.   In all cases, the BGOOD data support the case for the contact term to dominate the cross section close to threshold, and it is anticipated that the improved statistical precision will further constrain effective Lagrangian fit parameters in future analyses.


\begin{figure}[h]
	\includegraphics[width=0.5\textwidth]{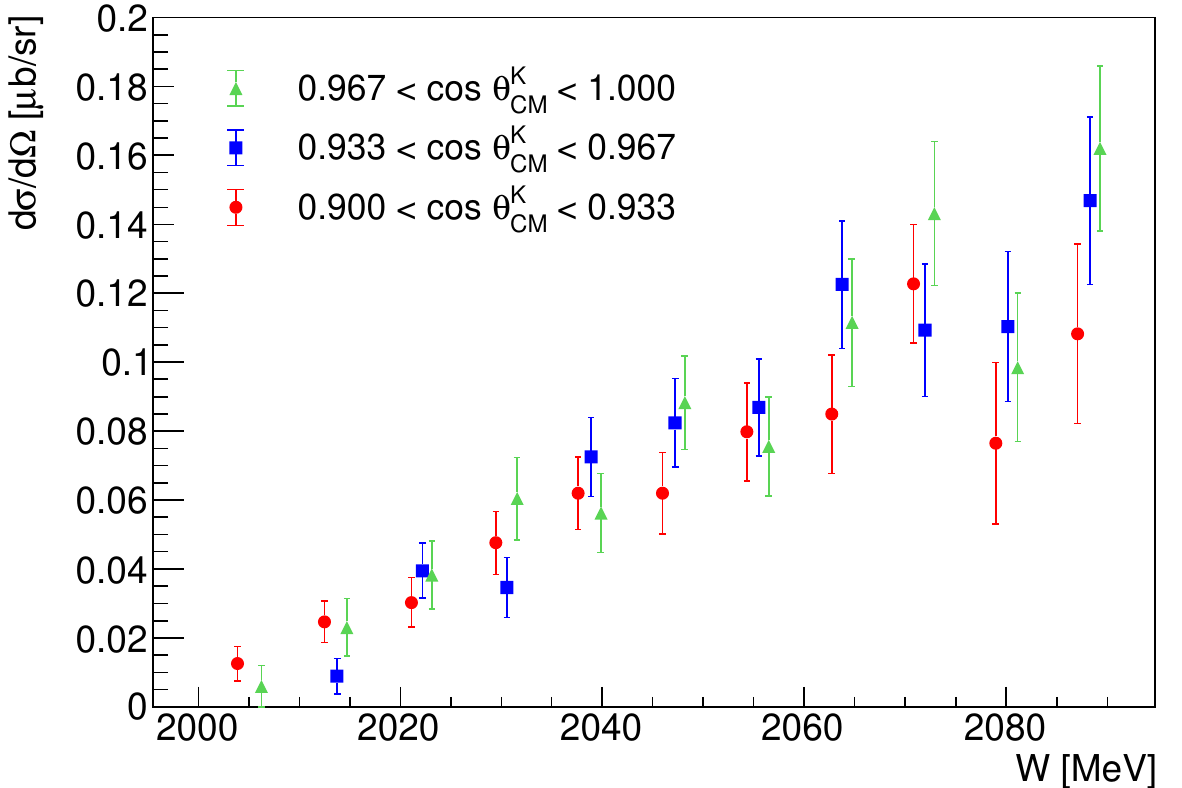}
	\caption{The differential cross section for fine intervals of  \kaonangle{}$> 0.9$.  Only the statistical error are shown.  The data points are staggered $-1$, 0 and +1\,MeV around the mean $W$ for clarity.}\label{fig:csfine}
\end{figure}

\begin{figure}[h]
	\includegraphics[width=0.5\textwidth]{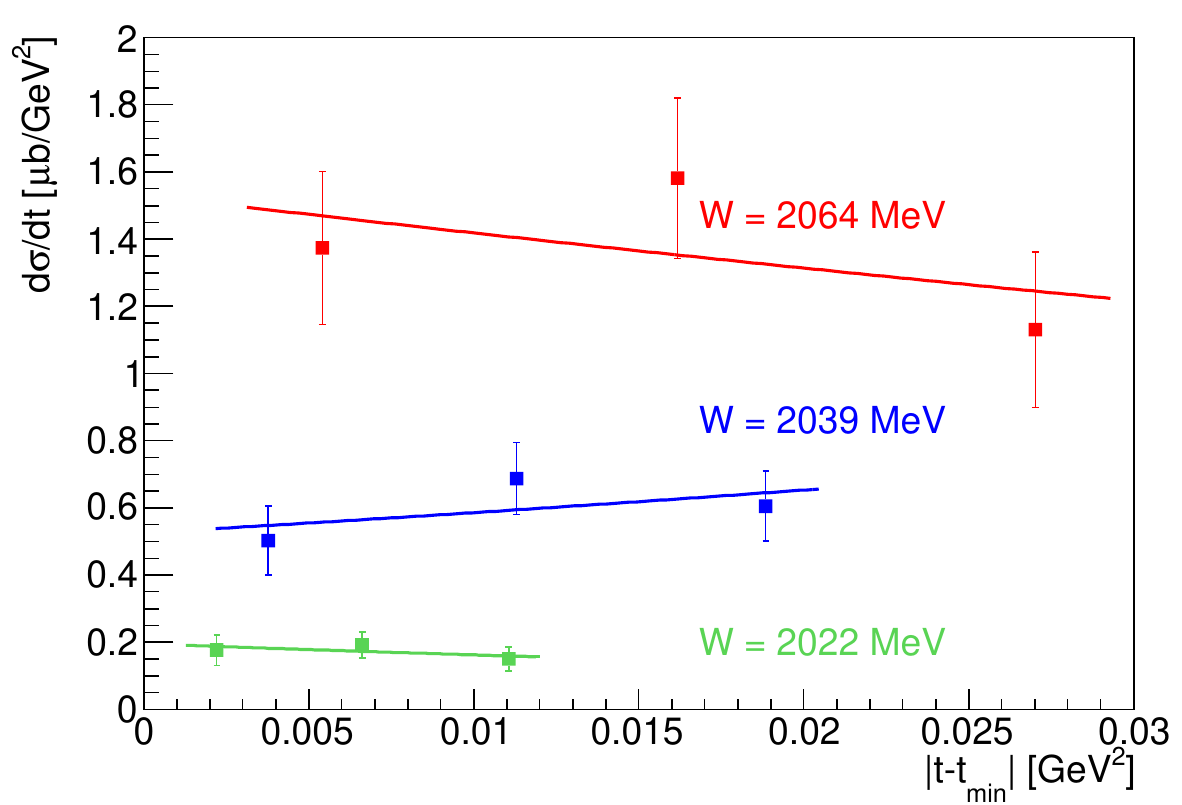}
	\caption{The differential cross section with respect to $t$ versus $|t - t_\mathrm{min}|$ for three example intervals of $W$ (only the statistical error shown).  Equation (1) is fitted to each $W$ interval, indicated by the colour coded lines.}\label{fig:csvst}
\end{figure}

The \kaonangle{} resolution of approximately 0.02 permits binning in finer \kaonangle{} intervals.  Fig.~\ref{fig:csfine} shows the same dataset in \kaonangle{} intervals of 0.033.  No significant change is observed across this \kaonangle{} range, which is in contrast, for example, to the $K^+\Sigma^0$ differential cross section which exhibits a cusp-like structure becoming more prominent in the most forward interval~\cite{ksigmapaper}.  This fine binning in \kaonangle{} provides an accurate means to extrapolate the data to the minimum possible momentum transfer to the $\Lambda(1520)$.  This is defined in terms of the Mandelstam variable, $t = (p_\gamma - p_K)^2$, where $p_\gamma$ and $p_K$ are the four-momenta of the photon beam and $K^+$ respectively.  To account for the distribution of $t$ within each two dimensional $W$ and \kaonangle{} interval, a generated distribution was used based on the measured differential cross section.  For each interval in $W$ and \kaonangle{}, the mean average value of $t$ was used as the central value, and the width was determined as $\sqrt{12}$\,RMS.  A fixed mass of the $\Lambda(1520)$ was assumed for the calculation in order to remove long tails to very high and very small $t$.  For each $W$ interval, the minimum momentum transfer, $t_\mathrm{min}$ was calculated, corresponding to \kaonangle = 1.  Fig.~\ref{fig:csvst} shows examples of the differential cross section with respect to $|t - t_\mathrm{min}|$ for three different $W$ intervals. 

The differential cross section at $t_\mathrm{min}$ was determined for each $W$ interval by fitting the function in Eq.~(1):

\begin{equation}
	\frac{\mathrm{d}\sigma}{\mathrm{d}t} = \frac{\mathrm{d}\sigma}{\mathrm{d}t}\Big|_{t=t_\mathrm{min}}e^{S|t-t_\mathrm{min}|}
\end{equation}\label{eq:fitfunction}

\begin{figure}[h]
	\includegraphics[width=0.5\textwidth]{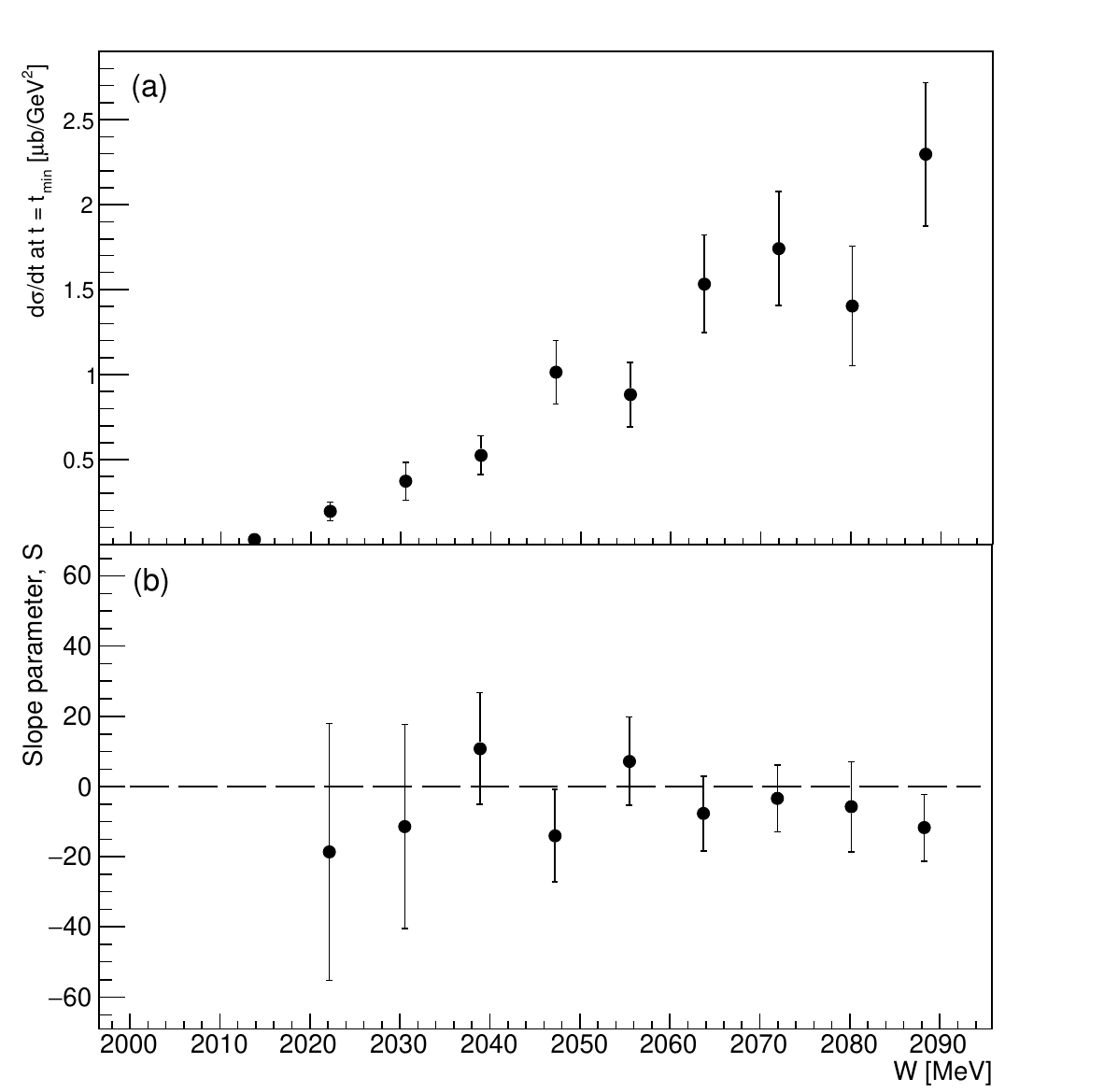}
	\caption{(a)  The differential cross section, d$\sigma$/d$t$ extrapolated to $t_\mathrm{min}$ versus $W$.  (b)  The slope parameter, $S$ (in Eq.~(1)) versus $W$.}\label{fig:cstattmin}
\end{figure}

\noindent Fig.~\ref{fig:cstattmin}(a) shows the extrapolated differential cross section at $t_\mathrm{min}$. 
  The slope parameter, $S$ is shown in Fig.~\ref{fig:cstattmin}(b).  Within limited statistical precision it appears to remain flat.  This is consistent with the contact term dominating near threshold, as was proposed by many  effective Lagrangian models~\cite{he12,nam05,nam07,nam10,toki08,xie10,xie14}.  If $K$ or $K^*$ $t$-channel exchange was dominant for example, the differential cross section could be expected to be more forward peaked, resulting in a negative value for $S$.

\section{Conclusions}

The differential cross section for $\gamma p \rightarrow K^+\Lambda(1520)$ was measured for \kaonangle{} $>$ 0.9 from threshold to a centre-of-mass energy of 2090\,MeV at the BGOOD experiment.  The resolution in both $W$ and \kaonangle{} enable a precise characterisation in this kinematic regime for the first time.  The data are consistent with the previous data of LEPS~\cite{kohri10} and effective Lagrangian models~\cite{he12,nam05,nam07,nam10,toki08,xie10,xie14}.  It is anticipated that the improved statistical precision will help constrain parameters in future phenomenological models, which can lead to an improved understanding of $t$-channel $K^*$, $u$-channel $\Lambda$ and $s$-channel $N^*$ contributions to the reaction mechanism.

\section*{Acknowledgements}

We thank the staff and shift-students of the ELSA accelerator for providing an excellent beam. 

We would like to thank Fei Huang and Neng-Chang Wei for discussion and providing tabulated data from Ref.~\cite{wie21}.

This work is supported by the Deutsche Forschungsgemeinschaft Project Numbers 388979758 and 405882627 and the Third Scientific Committee of the INFN.   This publication is part of a project that has received funding from the European Union’s Horizon 2020 research and innovation programme under grant agreement STRONG–2020 No.~824093. 

%
%
%


\setcounter{tocdepth}{3}

\end{document}